\documentclass{aa}
 \input epsf
 \begin{document}

\title{Ringing effects reduction by improved deconvolution algorithm\thanks{Based on observations obtained at the
Canada-France-Hawa\"{\i} Telescope (CFHT) which is operated by the
National Research Council of Canada (NRCC), the Institut des Sciences
de l'Univers (INSU) of the Centre National de la Recherche Scientifique (CNRS) and the University of  Hawa\"{\i} (UH), and on 
observations made with the NASA/ESA Hubble Space Telescope, obtained at 
the Space Telescope Science Institute, which is operated by the 
Association of Universities for Research in Astronomy, Inc., under NASA 
contract NAS 5-26555.}
 \subtitle{Application to A370 CFHT image of gravitational arcs}
}
\author{ Roche M.\inst{1,2} \and Bracco C.\inst{1} \and Aime C.\inst{1} 
 \and Lant{\'e}ri H.\inst{1} \and Mellier Y.\inst{3,4}}
\institute{UMR 6525 Astrophysique, Universit{\'e} de Nice Sophia Antipolis Parc Valrose, F-06108 Nice cedex 2          
\and
Institut Fresnel, UMR 6133, Ecole Nationale Sup{\'e}rieure de Physique de Marseille, Domaine Universitaire de Saint J{\'e}r{\^o}me, F-13397 Marseille cedex 20\\
 \email{muriel.roche@fresnel.fr}
\and
             Institut Astrophysique de Paris, UMR 7095, 98 bis Boulevard Arago, F-75014 PARIS
\and
Observatoire de Paris, LERMA, UMR 8112, 61 avenue de l'Observatoire, F-75014 Paris.
}
              \offprints{Roche M.}
 \date{Received xxx; accepted yyy}

\abstract{We develop a self-consistent automatic procedure to
 restore informations from astronomical observations.  It relies on
 both a new deconvolution algorithm called LBCA (Lower Bound Constraint Algorithm) and
 the use of the Wiener filter. In order to explore its 
  scientific potential for strong and weak gravitational lensing, 
  we  process a CFHT image of the galaxies cluster Abell 370 which
exhibits spectacular strong gravitational lensing effects. A high
quality restoration is here of particular interest to map the dark
matter within the cluster. We show that the LBCA turns out specially 
  efficient to reduce
ringing effects introduced by classical deconvolution
 algorithms in images with a high background. The method allows us to 
  make a blind detection of the radial arc and  to recover morphological 
  properties similar to those observed from HST data.  We also show that the Wiener 
 filter
 is suitable to stop the iterative process before noise amplification,
  using only the unrestored data .
  }

\titlerunning{Ringing effects reduction...}%
\authorrunning{Roche.M et al.}%

\maketitle

 \section{Introduction}
Despite its oustanding image quality, the small field of view of the Hubble
Space Telescope (HST) still hampers its use for deep  surveys covering angular 
  size
beyond a degree scale. Wide field surveys, like those used for gathering large
sample of lensing clusters or for cosmic shear, are therefore 
 a specific territory  for 
panoramic CCD cameras, like Megacam at CFHT [\cite{boul00}] or Omegacam at
ESO [\cite{vale01}]. However, the intrinsic limitation of ground-based
telescopes produced by atmospheric seeing  puts severe bounds on the
detection limits of these surveys and on the lowest gravitational distortion
amplitude one can measure with these cameras. Image degradation 
dilutes the light from small faint galaxies below the limiting threshold, blurs image details and increases the uncertainties on shape measurement of
lensed galaxies. Both arc detection and cosmic shear signal are therefore
altered by the seeing. \\
Improving image quality from ground based telescopes is
therefore an important technical goal that may have a significant scientific 
 impact when surveys are pushed to the limits. 
 In principle, image deconvolution can both improve the image quality
and  enhance the flux emitted by low surface brightness galaxies.
Unfortunately, because deconvolution is a slow process and often produces
unwanted artifacts, like ringing, it cannot be easily used on wide field ground
based images. Furthermore, in the case of very large data sets from 
panoramic cameras, objective convergence criteria must be defined and applied
automatically to images. These technical limitations turn out to be challenging
if one envisions a massive image deconvolution of surveys. 
\\
Although it is not yet applicable to large ground based data sets, we 
explore new image deconvolution techniques that could be used in the future.
More precisely, we compare the performances of the Richardson-Lucy
 (RL) algorithm
with a modified version called the Lower Bound Constraint Algorithm (LBCA) that
has been developped in [\cite{lant01}]. 
This algorithm, as well as the RL algorithm, shows noise amplification when the
iteration number increases too much. A procedure to control automatically the
iteration has then been implemented and validated on real data. It 
  uses a  HST image as  a reference to
 stop the deconvolution process when a given distance between the HST image and 
  the restored image is minimum. We choose to minimize an Euclidean distance
criterion. This minimization technique also
   enables us to check the quality of the restoration. 
\\
Although the comparison with HST data turns out to be a successful way to 
  test the method, it is unpractical for most images. 
 Operating cosmological surveys will provide a huge amount of data that must be
automatically processed, without any reference images. So, we 
 generalized our empirical convergence criterion using the HST images
   to develop a
systematic procedure to stop the deconvolution algorithm. This technique is
based on the Wiener filtering and relies on the information contained in the
data only. 
\\
In this first paper, we applied the deconvolution on CFHT images of the famous
giant arc in Abell 370 and compare the result to HST data. Both the CFHT and
the HST images are described in section 2. We  determine the Point Spread
Function (PSF) in the CFHT image in section 3 and apply the deconvolution
techniques in section 4. We show that the LBCA must be preferred to the
standard RL algorithm.  In section 5, we use the technique based on the Wiener
filter to stop the LBCA iterations without the need of any reference image.  We
show that the quality of the restored image then obtained with this independent
 procedure is satisfactory. We sum up our conclusions  in section 6.

\section{The data}

Abell 370 is the most distant cluster of galaxies in the Abell catalogue, at a
redshift z = 0.374 [\cite{sara82}]. The cluster structure is dominated by
two giant elliptical galaxies, identified as 
  $\#20$ and $\#35$ from the notations in [\cite{souc87b}]\footnote{We use thereafter the notations of  the
  [\cite{souc87b}] and [\cite{mell98}] to refer to the
details in our images}. The image of Abell 370 exhibits a giant arc extending
over 60 arc-second wide discovered by Soucail et al. [\cite{souc87}] 
  and [\cite{lynd86}]. This arc has been identified as an extremely distorted image
of a background galaxy at z=0.724 [\cite{souc88}], thus bringing
the evidence of a cosmological gravitational lensing effect. The arc is split
into five distinct regions  $\#b$, $\#c$, $\#g$, $\#37$
 and $\#62$ (see Fig. 6. in the present paper) and
shows an intensity breaking between the galaxies $\#37$ and 
 $\#62$. The galaxies $\#37$  
($z_{\#37}=0.37$), $\#b$ ($z_{\#b}=0.363$) and $\#c$
  ($z_{\#c}=0.373$) are superimposed to the arc. On the
contrary, the object $\#62$ belongs to it. 
The detection of a weak radial arclet in a HST/WFPC1 image (noted R) has been
reported in [\cite{smai96}]. Its existence was confirmed in [\cite{beze99}] who argued  its redshift should be about 1.3. The detection
of arclets, as well as the determination of sub-structures in the giant arc,
are important to constrain the mass profile of Abell 370 as well as to
  scale its absolute mass. 

\subsection{CFHT observations of A370}
The A370 image was obtained at the 3.60m telescope of the CFHT
(Canada France Hawa\"{\i} Telescope) observatory in November 1991 
 (see [\cite{kneib93}] for details).
   The FOCAM (Faint Object CAMera) imager
installed on the prime focus of the telescope uses the filter
$\#$1808  centered on 832 nm with a width of 196 nm. This image is
integrated over 1800 seconds and corrected with standard CCD frame
packages. The field extends over about 6x6 arc-minutes ('), with a 
  sampling of 0.206 arc-second per pixel. The
objects of specific interest correspond to the strong lensing
effects observed close to the central galaxy ($\#$ 35 on Fig.
\ref{compaCHFHHST} a).  Thus we extract a sub-image extending over
0.6' $\times$ 0.6' approximately centered on this galaxy (image
196$\times$196 pixels).

\subsection{HST observations of A370}
We also have an A370 image with 622 seconds of integration time
obtained with the HST in a wavelength domain close to the CFHT
observations (F675W filter centered on $\lambda$=673,5 nm and
width of 88,9 nm). The WFPC2 (Wide Field Planetary Camera 2)
gives a wide field about  150 $\times$ 150 arcsec ($"$).  We have extracted a region of about  34 $\times$ 34 $"$ (0,1"/pixel) containing the region of the giant arc. The high spatial
resolution of the HST allows to observe details invisible in the
CFHT image as sharp breakings in the giant arc and the presence
of the radial arc $R$.  We make the assumption that the  HST image can be used as a reference for comparison with the restored image, i.e. we neglect any alteration induced from the HST instrument. We cannot exclude differences between the two original astronomical fields, because they are seen through different filters, but these differences are probably negligible for the study undergone here. This assumption is also supported by a mere visual inspection of the two images.

\subsection{Comparison between HST and CFHT observations}
\begin{figure}[t]
\centerline{\epsfxsize=7cm\epsfbox{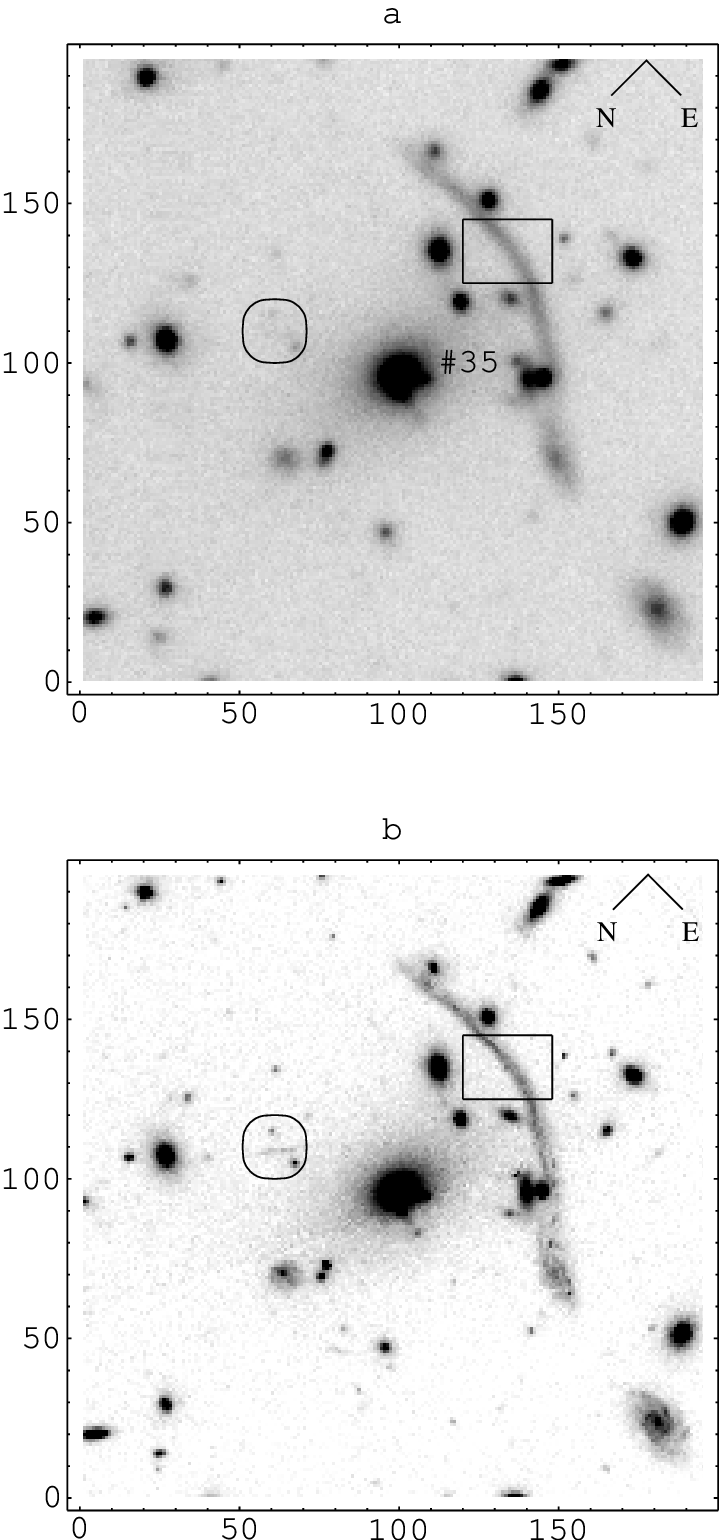}}
  \caption[]
  {a) Extraction of a sub-image ($196 \times 196$ pixels) of Abell 370 observed with the CFHT. The circle
  represents the region where the radial arc $R$ must be located.\\The rectangular box ($20 \times 28$ pixels) represents the mask used for the comparison of the two images in the
  case of the arc reconstruction.\\
  b) Extraction of a sub-image ($196 \times 196$ pixels) of Abell 370 image obtained by applying the linear transformation (\ref{plane}) on the HST image. The circle
  identifies the radial arc  $R$. }\label{compaCHFHHST}
\vspace{0.2cm}
\end{figure}

 The direct comparison
between CFHT and HST images requires a linear transformation of
the HST image since both orientation and spatial sampling are
different. Let $(x',y')$ and $(x,y)$ be the spatial
coordinates of a star photocenter in the HST and CFHT images
respectively. The linear transformation between $(x',y')$ and
$(x,y)$ may be written as:
\begin{equation}\label{plane}
  \left \{ \begin{array}{l} x'=a.x + b.y + c \\
 y'=d.x + e.y + f \end{array} \right.
\end{equation}
where $a$, $b$, $c$, $d$, $e$ and $f$ are unknown scalars. We then compute the linear transformation of the HST image with a bilinear interpolation. Solving the system (\ref{plane}) for a set of stars both identified in the CFHT and the HST 196$\times$196 sub-images, allows an accurate determination of the scalars  $a$, $b$, $c$, $d$, $e$ and
$f$. The comparison between the CFHT image and the transformed HST one is presented in Fig. \ref{compaCHFHHST}.\\
The transformation (\ref{plane}) may be interpreted as a
combination of a 110$^\circ$ rotation followed by a 2.07 pixels
dilatation from the center of the galaxy $\#$35 and a translation.

\section{The Point Spread Function (PSF)}
Ground-based observations suffer degradations due to the
transmission of light by the atmosphere and the optics. The images
obtained are then blurred by the system
atmosphere-instrumentation. The field of the CFHT image extends
over a few arc-minutes and corresponds to a long exposure time. In
these conditions we can make the assumption that the degradation
due to the atmosphere is space invariant in the image plane. Such
an assumption is also safe for the instrumental distortions. The
relation between the noiseless image $\tilde{y}(r)$ and the object
$x(\alpha)$ is then a convolution one:
\begin{equation}\label{fred}
  \tilde{y}(r) = \int_{-\infty}^{+\infty}h(r-\alpha)x(\alpha)d\alpha   =
   h(r)\odot x(r)
\end{equation}
 where $r$ is the spatial
coordinates at the telescope focus and $h$ the kernel of the integral equation (\ref{fred}) is the PSF of the system atmosphere-telescope.\\
The discretization of equation (\ref{fred}) leads to the
 matrix relation:
\begin{equation}
 \tilde{y}=Hx
 \end{equation}
where $\tilde{y}$ is a vector corresponding to the noiseless
image, $x$ a vector corresponding to the object and $H$ a matrix
representing the convolution effects of the PSF. In real cases, the observations $y$ are a
noise corrupted version of
$\tilde{y}$. In all what follows, we assume a Poisson noise process.\\
 The restoration technique we propose in
this paper are then based on the deconvolution of the noisy image
$y$ by an estimated PSF in order to reconstruct the object $x$
in the best conditions.

The first step consists in estimating the PSF from the stars
(unresolved objects) in the full CFHT image.

\subsection{Stars selection} We first identify the stars by using SExtractor [\cite{bert96}]. This software extracts
the objects in the field and gives characteristic parameters such
as the centro$\ddot{\i}$d position, the Full Width at Half Maximum
(FWHM) in the light distribution and the magnitude of the object.
An object is defined as a number of connected pixels (fixed by
the user) above a given threshold (1.5 $\sigma$ in our study,
where $\sigma$ corresponds to the standard deviation of the image
background assuming a Gaussian distribution). The star
identification is then achieved by representing on a diagram the
magnitude of each object versus the FWHM. The stars gather in a
region with a FWHM approximatively constant (between 0.6" and 0,68")
independently of the magnitude. A refined selection is carried out
by only preserving the stars whose magnitude is greater than a
lower bound to avoid saturated stars images and smaller than a
higher one to avoid galaxy contamination. Each star image is
extracted over a 32 $\times$32 pixels frame. After dropping
images exhibiting neighbors we are left with a sample of 23
individual stars images.

\subsection{Estimation of the PSF} 
\begin{figure}
\centerline{\epsfxsize=5cm\epsfbox{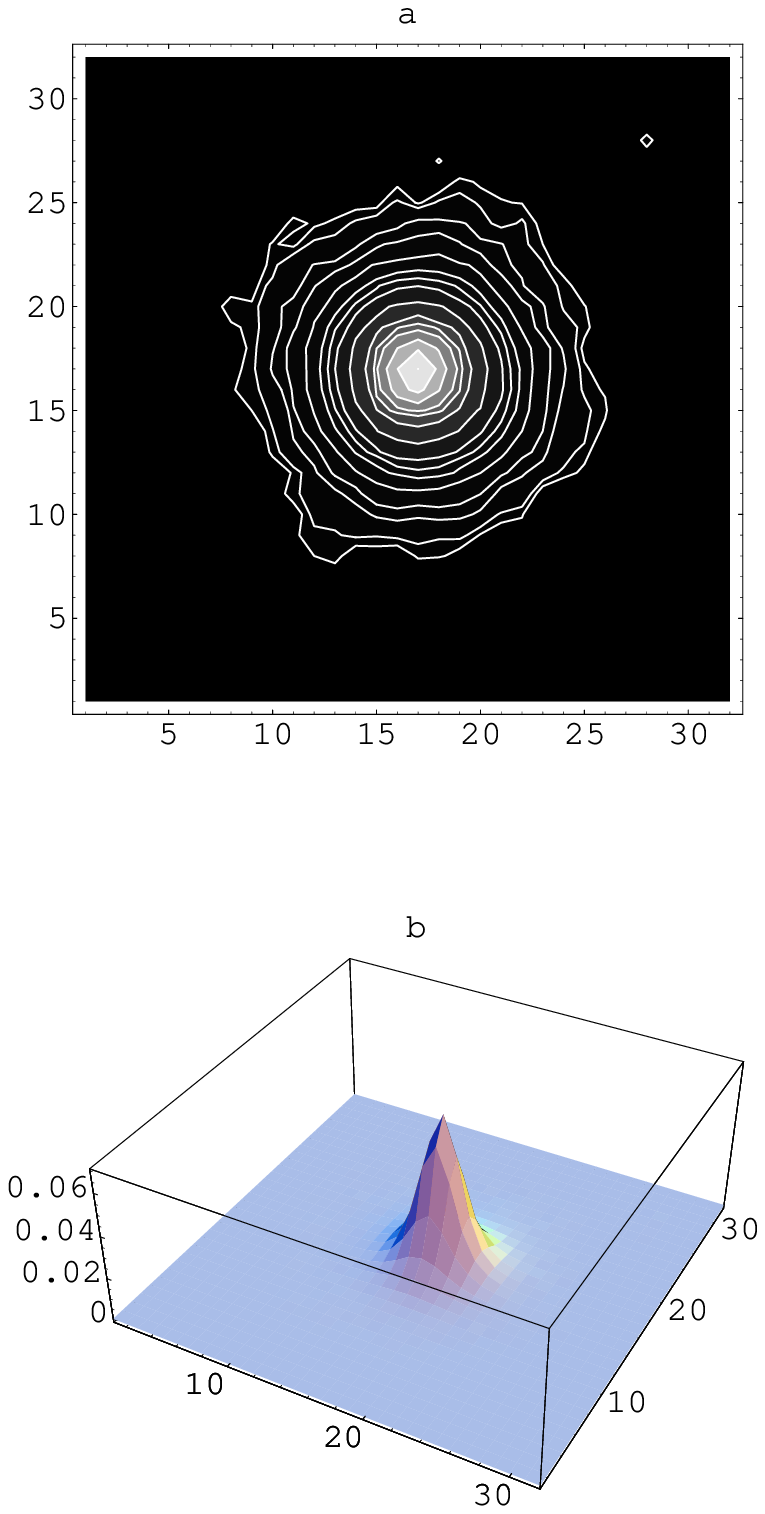}}
  \vspace{0.2cm}\caption[]
  {Normalized PSF of the A370 image. \\
  a) Contours Plot. The contours represent successively 100, 80, 60, 40, 20, 10, 5, 3.5, 2.5, 1.5, 0.8,
  0.5, 0.2, 0.1 $\%$ of the image maximum. \\ b) 3D Representation.}
  \label{psfaniso}
\end{figure}

The second step
consists in adding the 23 stars to synthesize the PSF. The
summation of these objects imposes the perfect superposition of
the photo-center of each star up to a fraction of pixel. This could
not be done directly as the light centro$\ddot{\i}$d is not
located on an full pixel. We have to estimate the vector shift
between each star $f_i(x,y)$ where $(x,y)$ are the spatial
coordinates and a reference one $f_*(x,y)$ chosen arbitrarily.
This is achieved by determining the argument of the inter-spectrum
between each couple of images. We first compute the Fourier
Transform (FT) for each star image noted $\widehat{f}_i(u,v)$ and
$\widehat{f}_*(u,v)$ for the reference, where $(u,v)$ are the
spatial frequencies. Let us denote $\xi_{i,1}$ and $\xi_{i,2}$ the
unknown shifts between $f_i(x,y)$ and $f_*(x,y)$ along the $x$ and
$y$ axes respectively. The argument $\varphi_i  (u,v)$ of the
inter-spectrum between $\widehat{f}_i(u,v)$ and
$\widehat{f}_*(u,v)$ is:
\begin{equation}
  \varphi_i (u,v)= - 2 \pi (u \xi_{i,1} + v \xi_{i,2})
\end{equation}
We fit $\varphi_i (u,v)$ by  a plane in the low frequencies range (where
the Signal to Noise Ratio (SNR) is better) to obtain $\xi_{i,1}$ and $\xi_{i,2}$. This
leads to an accurate determination of the shift vector in the
direct space. Each star is then translated in this latter space to the position of the
reference by using a bilinear interpolation. Since the background, noted $back$, changes slightly in
the field, we substract its mean specific value in each star
image. We finally compute a SNR for each
image, defined as:

\begin{equation}
  SNR = \frac{(Max_*)^2}{Max_* + back}
\end{equation}
where $(Max_*)$ represents the maximum intensity of the star.  The images
are then weighted by their SNR and summed up. The result is
normalized to yield the PSF in the field. This PSF shows a rather
axial symmetry with a slight anisotropy (Figure \ref{psfaniso}).
The PSF is then put into a 196$\times$196 frame as the deconvolution process requires. We denote by $H$ the matrix
corresponding to the effects of the PSF.

\section{Deconvolution of the A370 CFHT image}

\subsection{Limitation of the standard Richardson-Lucy algorithm: ringing effects}
We first deconvolve the A370 image with the classical
Richardson-Lucy algorithm [\cite{lucy74}, \cite{rich72}] and the
PSF previously estimated:

\begin{equation}
x^{(k+1)} = x^{(k)}.H^T\frac{y}{Hx^{(k)}}
\end{equation}
where $x^{(k)}$ represents the reconstructed object at the
iteration $k$, $y$ is the observation, $H$ is the PSF matrix
 and $H^T$ its transposed.
\\

\begin{figure}[t]
\centerline{\epsfxsize=6cm\epsfbox{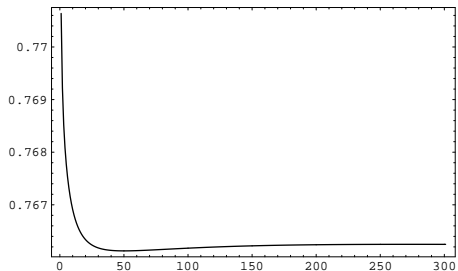}}
  \caption[]
  {Relative error curve $R_k$ between the CFHT reconstructed image and the HST one as a function of the iteration number for RL.
  The images used are those of the figure \ref{compaCHFHHST}, the PSF is the one of figure
  \ref{psfaniso} and the mask used correspond to the rectangular box ($20 \times 28$ pixels).}\label{erreur50}
\end{figure}

\begin{figure}[t]
\centerline{\epsfxsize=7cm\epsfbox{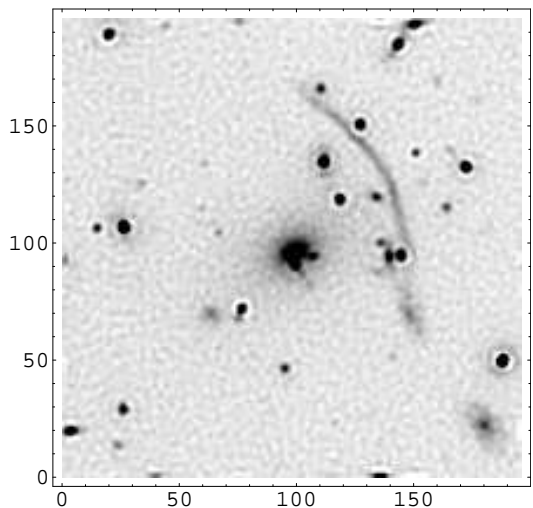}}
  \caption[]
  {Abell 370 reconstructed image by RL at the 50th iteration.
  The image used is the one of figure \ref{compaCHFHHST} a), the PSF is the one of figure
  \ref{psfaniso} and the mask used correspond to the rectangular box ($20 \times 28$ pixels).}\label{recons50}
\end{figure}

 To stop the iterations of the algorithm we define an error
 criterion between the image reconstructed at
iteration $k$ (denoted $x^{(k)}$) and a reference image. In this
section we use the HST image (denoted $x^{HST}$) as the reference.
We then minimize a criterion $R_k$ based upon a relative Euclidean
distance between $x^{(k)}$ and $x^{HST}$:
\begin{equation}\label{ro}
 R_k =
\frac{\parallel x^{HST} - x^{(k)}
\parallel }{\parallel x^{HST} \parallel}
\end{equation}
where $ \parallel .
\parallel$ represents the Euclidean
distance.\\
 The objects in the field are rather different
(giant arc, galaxies, stars...) and would not require exactly the
same iteration number to be reconstructed at the
best. The criterion $R_k$ must then be computed preferably over
the area to be preferentially restored. We focus here on the
region of the giant gravitational arc and thus minimize $R_k$
(see Figure \ref{erreur50}) within the rectangular box indicated in Figure
\ref{compaCHFHHST}. The best reconstructed image obtained at iteration $k=50$ is represented in Figure \ref{recons50}.
Its background shows a granular structure typically of the size of
the PSF. An early ringing effect [\cite{cao99}, \cite{lucy94},
\cite{lage91}] appears around the brightest point-like objects. It
is both caused by the important background in the CFHT image and
the discontinuities in the FT due to bright point-like objects  [\cite{whit93}].
These phenomena  prevent any accurate measurement on the restored
image and render necessary the use of a modified version of the
algorithm to remove the oscillations. It is achieved by taking
into account the background in the image and by introducing it as a
lower bound constraint in the deconvolution algorithm itself.
Hence, the reconstruction is not allowed to take values less than
the background.

\subsection{Amelioration: Lower Bound Constraint Algorithm (LBCA)}
 The LBCA  is developed from a method
proposed in [\cite{lant01}] and studied in detail in
[\cite{roch01}, \cite{lant02}]. The method is based on the
minimization of a convex function under lower bound constraint
(denoted $m$). It consists in writing that at the optimum, the
Kuhn-Tucker conditions [\cite{kuhn51}] are fulfilled. We then
write the algorithm under a modified gradient form using the
 successive substitution method [\cite{hild74}]. After simple algebra, this leads  in the
non-relaxed case to the following multiplicative expression of the
LBCA:
\begin{equation}\label{rlinf}
 x_{i}^{(k+1)} = m_{i} +  (x^{(k)}_{i} - m_{i}) \left[H^T
\frac{y}{(Hx^{(k)})} \right]_{i}
\end{equation}
where $i$ is the current pixel in the image.  A similar algorithm has been proposed by [\cite{snyd93}] and by [\cite{nune93}] with however a slightly different formulation. A comparison between the two approaches is given in the Appendix. Note that when $m=0$ whatever the iteration, the RL algorithm is recovered.\\
 The lower bound
$m$ can be constant or variable over the image. The algorithm
implementation is just as simple in both cases. The difficulty
consists in obtaining an accurate background map specially in the
region of bright structures. This estimation is out of the scope
of this paper. We have then chosen to take for $m$ a constant
value overall the image estimated as the mean background
in the global normalized CFHT image.\\

\begin{figure}[t]
\centerline{\epsfxsize=6cm\epsfbox{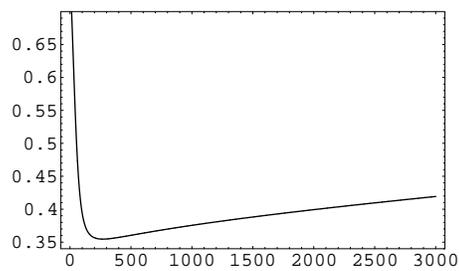}}
  \caption[]
  {Relative error curve $R_k$ between the CFHT reconstructed image and the HST one as a function of the iteration number for LBA.
  The images used are those of the figure \ref{compaCHFHHST}, the PSF is the one of figure
  \ref{psfaniso} and the mask used correspond to the rectangular box ($20 \times 28$ pixels).}\label{erreur269}
\end{figure}

\begin{figure}[t]
\centerline{\epsfxsize=7cm\epsfbox{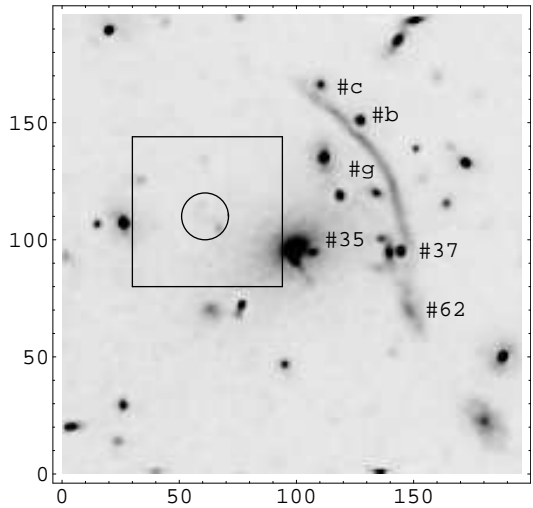}}
  \caption[]
  {Best reconstructed image with LBCA using the HST reference image at the iteration 269.}\label{recons269}
\end{figure}

 We still stop the LBCA by minimizing the $R_k$ criterion (\ref{ro}).
 We evaluate $R_k$ within the rectangular box of Figure
\ref{compaCHFHHST} as in the previous section. The corresponding
$R_k$ curve is represented on Figure \ref{erreur269}. The best
restoration for this region of the giant arc is obtained for the
iteration 269 (Figure \ref{recons269}). We see from Figure
\ref{erreur269} that a range of about 50 iterations around this
value would yield reconstructions of similar quality.

\begin{figure}[t]
\centerline{\epsfxsize=5cm\epsfbox{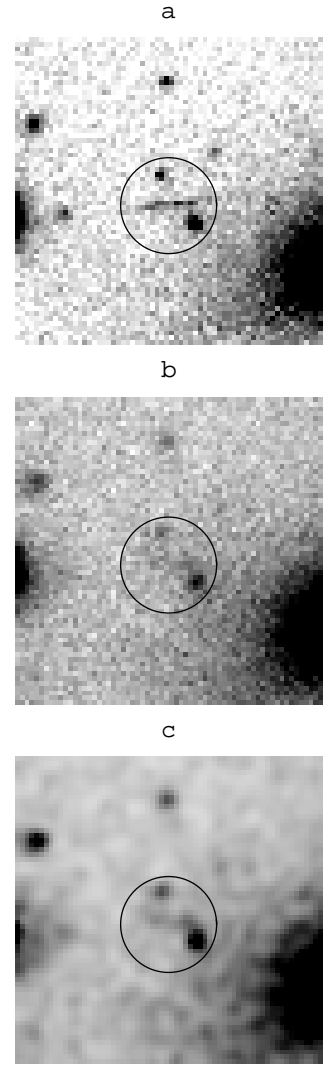}}
  \caption[]
  {Zoom on the radial arc R. HST image (a), raw CFHT image (b) and deconvolved CFHT image (c). }\label{Rarc269}
\end{figure}
 The reconstructed image with the LBCA shows an important
reduction of both granularity and  ringing effects.\\
 We use a criterion derived from $R_k$ to measure the amelioration brought by the deconvolution.
 We compute a quantity $D$ defined by:
\begin{equation}\label{d}
  D = 100 \times \frac{R_y - R_{opt}}{R_y}
\end{equation}
where $R_y$ and $R_{opt}$ are defined as:
\begin{equation}\label{ressem}
R_y = \frac{\parallel x^{HST} - y\parallel}{\parallel x^{HST}
\parallel}\; \;  \; , \; \; \; R_{opt} = \frac{\parallel x^{HST} - x^{(k_{opt})} \parallel}{\parallel x^{HST}
\parallel}
\end{equation}
$R_y$ and $R_{opt}$ measure the relative Euclidean distances
between the HST reference $x^{HST}$ and the CFHT image before
deconvolution $y$, and between $x^{HST}$ and the optimal
restoration $x^{(k_{opt})}$. $D$ gives the averaged amelioration
in percent with respect to the Euclidean measure.
 It raises up to 53$\%$ for the
giant arc. We have also carried the minimization of $R_k$ for the
full 196$\times$196 image and obtained a 47$\%$ amelioration with
respect to the $D$ criterion. This means that the image obtained
after deconvolution with the LBCA is closer to the HST image by
roughly a factor 2 than the image before deconvolution. No
significant amelioration is brought by the RL algorithm with this criterion.\\

 This study shows a great
amelioration carried out by the simple introduction of the lower
bound constraint. \\ The deconvolution allows
in particular to restore severals structures in the giant arc ($\#$b,
$\#$c, $\#$g, $\#$37, $\#$62), and the breaking between $\#$37 and
$\#$62.
Moreover, the reconstructed image evidences a radial
gravitational arc (circle in Figure \ref{recons269} and
\ref{Rarc269}.c). This arc is clearly visible in the HST image (Figure \ref{Rarc269}.a) but does not appear in the raw CFHT data  (Figure \ref{Rarc269}.b). \\Ringing residuals, while attenuated, are still present around high intensities objects. These drawbacks are due to an incorrect estimation of the background in these regions. The reconstruction can be improved by
estimating a variable background.

We emphasize that the LBCA process takes the same computation time as the RL algorithm.\\

\section{Wiener filtering for stopping iterations}
\begin{figure}[t]
\centerline{\epsfxsize=6cm\epsfbox{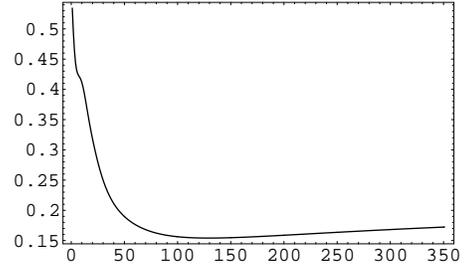}}
  \caption[]
  {Relative error curve $E_w(k)$ between the modulus of the FT of the reconstructed image at
iteration $k$ (Abs$[\widehat{x}^{(k)} (u,v)]$) and the FT of the
Wiener reconstructed image (Abs$[\widehat{x}_w(u,v)]$). }\label{ew}
\end{figure}

\begin{figure}[t]
\centerline{\epsfxsize=7cm\epsfbox{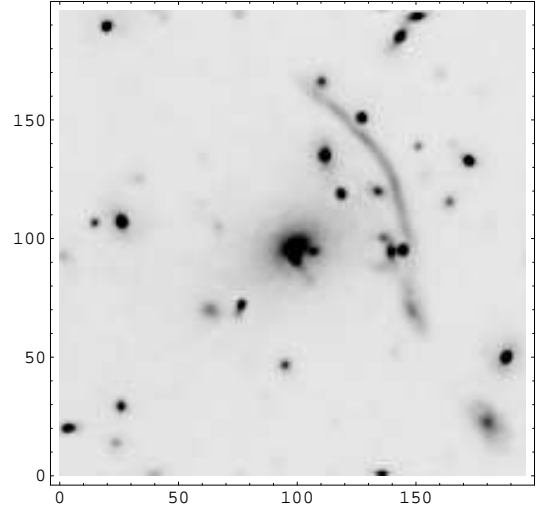}}
  \caption[]
  {Best reconstructed image with LBCA using a criterion based on the Wiener filter at the 129th iteration.}\label{recons129}
\end{figure}

We have used in the previous section a HST image to stop the iterations of
 the deconvolution. Usually, such a reference image is not
 available. Further, the objective is precisely to restore information
 from the sole ground-based observations without any external
 help. So it is necessary to develop a self-consistent method relying
 on the ground-based  data only.\\
  The technique we used is the one proposed by Lant{\'e}ri et al. [\cite{lant299}]. It makes use of a comparison of  the modulus of the Fourier transform of the deconvolved image at the iteration $k$ with the modulus of the Fourier transform given by a Wiener filtering. 
For clarity, let us briefly recall some well-known results on Wiener filtering. The Wiener filter $W(u,v)$ is a zero-phase filter that prevents the amplification of the noise if a raw inverse filter technique is used. Denoting  $r$  the angular coordinates and  $u$ and $v$ the 2D spatial frequencies, $\widehat{y}(u,v)$ and $\widehat{h}(u,v)$ the FT of the observation $y(r)$ and the PSF $h(r)$, the inverse Wiener filtered image transform $\widehat{x}_w(u,v)$ may be written as  [\cite{brau71}]:
\begin{equation}\label{xw}
\widehat{x}_w(u,v)=\frac{\widehat{y}(u,v)}{\widehat{h}(u,v)} W(u,v)
\end{equation}
where $W(u,v)$ is defined as:
\begin{equation}\label{wien}
W(u,v)=\frac{P_{Hx}(u,v)}{P_{Hx}(u,v)+ P_{n}(u,v)}
\end{equation}
 The quantities $P_{Hx}(u,v)$ and $P_{n}(u,v)$ are the power spectra
 of the noiseless image and of the noise itself respectively.
Estimating these quantities is the main difficulty in the implementation of a Wiener filter. Only approximated expressions can be worked out. Assuming that the signal and noise are statistically independent,  $P_{n}(u,v)$ can be taken as a constant. Its value may be estimated in the very high frequencies of  $\widehat{y}(u,v)$, where no astronomical signal is expected. Estimating $P_{Hx}(u,v)$ is more difficult since it implies an a priori knowledge of the object FT. In the present work, we have assumed that $P_{Hx}(u,v)$ could be approximated by $P_{h}(u,v)$, the power spectrum of the telescope-atmosphere PSF. This tends to overestimate  $W(u,v)$.
	So, after this procedure, equations \ref{xw} and \ref{wien} permit to have an estimate of $\widehat{x}_w(u,v)$. Following the work of Lant{\'e}ri et al. [\cite{lant299}], we use only its modulus Abs$[\widehat{x}_w(u,v)]$ to define a stopping criterion for the LBC algorithm. 
	At each iteration $k$, we compute the euclidean distance  $E_w(k)$ of the form : 
\begin{equation}
E_w(k)=  \frac{ \parallel  \rm{Abs}[\widehat{x}^{(k)} (u,v)]-
\rm{Abs}[\widehat{x}_w(u,v)]\parallel }{ \parallel  \rm{Abs}[\widehat{x}_w(u,v)]
\parallel}
\end{equation}
The raw application of this criterion fails to give a satisfactory iteration stop. A much better solution was found in using only in this comparison a range of intermediate frequencies. Indeed it was found necessary to suppress the very low frequencies that have a too important weight in the calculus. The comparison must also exclude the highest spatial frequencies that are suppressed by the Wiener filter used. The overall effect is to apply an annular mask in the frequencies plane.\\
Both Abs$[\widehat{x}^{(k)} (u,v)]$ and Abs$[\widehat{x}_w(u,v)]$
are normalized inside this mask for comparison. The
distance $E_w(k)$ is represented as a function of the iteration number $k$
on Figure \ref{ew}.
The best reconstruction is obtained for the iteration k = 129
(Figure \ref{recons129}).
The amelioration brought by the deconvolution
estimated from the $D$ criterion (\ref{d}) reaches 48$\%$. It is close to the 53$\%$ amelioration obtained in section 4.2 using the HST reference. It
shows that independently of a reference HST image, a rather
satisfying approximate solution may be found by using a stopping
criterion based on the data themselves. This procedure can be
fully and easily automated.

\section{Conclusion}

We have used a modified version of the RL algorithm, called LBCA, to deconvolve a CFHT image of Abell 370. The LBCA introduces a lower-bound constraint which 
 prevents the reconstructed image to take values below this bound. It allows 
  to reduce considerably the ringing effect that appears around bright objects 
 when classical algorithm are used (in particular RL). In the present paper, 
 the lower bound is taken equal to an estimation of the mean sky background 
  over the whole CFHT 196$\times$196 image.
\\
We have used at first an HST image of A370 as a reference to stop the iteration of the algorithm and to evaluate the amelioration brought by the 
  deconvolution. An Euclidean distance criterion is then minimized between 
  the reconstructed image at a given iteration and the HST reference to yield 
  the best reconstruction. 
   The LBCA reconstruction of A370 is twice as close to the HST image 
  than the raw CFHT data, while no amelioration is brought by using the RL 
 algorithm.  Remarkably, our iterative process has permitted to detect 
  blindly the radial arc evidenced earlier on  HST image of  A370 and  
    to recover its morphological properties. This is
   an encouraging demonstration of its efficiency, and an interesting  
 example of practical application. It also  emphasizes the interest of the 
  LBCA to restore images with a high background.
\\
In a more general case where no HST image is available, the 
  convergence and stopping rules of the algorithm must rely on the information 
 contained in the data themselves. We have thus developed a technique using 
  the Wiener filter. It proceeds from an estimation of the noise in the CFHT image and the evaluation of the Power spectrum of the PSF. This method is based on the minimization of  an Euclidean distance between the FT modulus of the LBCA reconstruction and the reconstructed image by the inverse Wiener filter. The best reconstructed image is close to the former one obtained with the HST as a reference.

The present study evidences two important results. First, it allows to improve the quality of an image with a high background, using a new algorithm as simple
and as fast as the RL one: the LBCA. Finally, the overall data processing involving the LBCA together with the Wiener filtering may be fully automated. Hence, it could be fruitfully used for the processing of huge amount of ground-based observations and particularly in the perspective of current or forthcoming wide field surveys. \\\\
\appendix
\section{}
 We show in this appendix, the similarities between the algorithm used in the present paper and the algorithm previously proposed by [\cite{snyd93}] and by [\cite{nune93}], to take into account the effect of the background.

In our algorithm written in the form \ref{rlinf}:
\begin{equation}
 x_{i}^{(k+1)} = m_{i} +  (x^{(k)}_{i} - m_{i}) \left[H^T
\frac{y}{(Hx^{(k)})} \right]_{i}
\end{equation}
$x$ is clearly the overall value of the solution, and we must have:
\begin{equation}
x_{i} \geq m_{i} \hspace{0.1cm}\forall i
 \end{equation}
Therefore, $x$ includes the background (the background $m$ is in the object space).
The model used is $\tilde{y}=Hx$ and the actual data $y$ is Poisson with mean $Hx$. Now, introducing $u=x-m$ in this algorithm, we obtain immediately:

\begin{equation}
 u_{i}^{(k+1)} = u^{(k)}_{i}  \left[H^T
\frac{y}{(H(u^{(k)}+m)} \right]_{i}
\end{equation}
This the form previously proposed by [\cite{snyd93}] and by [\cite{nune93}].\\
Here the model used is $\tilde{y}=H(u+m)$, $u$ is the part of the signal over the background $m$, and $u_i \geq 0$ $\forall i$. The background is again in the object space and the actual data $y$ is Poisson with mean $H(u+m)$.\\
Except these small conceptual differences, both algorithms are similar and leads to the same results, however another important point must be underlined.
Because of the convolution, the mean $H(u+m)$ can also be written $Hu + m$ so the background $m$ appears in the data space). However it does not mean that we can subtract the background $m$ from the data $y$ for obtaining modified data $y'=y-m$; indeed if $y$ is Poisson with mean $Hu + m$ , then $y'$ is not Poisson of mean $Hu$ ($y'$ might also be negative).\\
Then, whatever the form of the algorithm, the estimated background is used in the algorithm [\cite{hani93}], but the data $y$ must left unchanged.


 \end{document}